\documentclass[a4paper,fleqn,usenatbib]{mnras}

\usepackage{newtxtext,newtxmath}
\usepackage[T1]{fontenc}
\usepackage{ae,aecompl}

\usepackage{graphicx}	% Including figure files
\usepackage{amsmath}	% Advanced maths commands
\usepackage{amssymb}	% Extra maths symbols
\usepackage{soul}
\title[The hidden satellites of quasars at $z \, = \, 6$]{The hidden satellites of massive galaxies and quasars at high-redshift}
\author[Costa, Rosdahl \& Kimm]{
Tiago Costa$^{1}$\thanks{E-mail: tcosta@mpa-garching.mpg.de},
Joakim Rosdahl$^{2}$
\& Taysun Kimm$^{3}$
\\
$^{1}$Max-Planck-Institut f\"ur Astrophysik, Karl-Schwarzschild-Stra{\ss}e 1, D-85748 Garching b. M\"unchen, Germany\\
$^{2}$CRAL, Universit\'e de Lyon I, CNRS UMR 5574, ENS-Lyon, 9 Avenue Charles Andr\'e, 69561, Saint-Genis-Laval, France\\
$^{3}$Department of Astronomy, Yonsei University, 50 Yonsei-ro, Seodaemun-gu, Seoul 03722, Republic of Korea
}

\date{Accepted 2019}

\pubyear{2019}

\begin{document}
\label{firstpage}
\pagerange{\pageref{firstpage}--\pageref{lastpage}}
\maketitle
% Abstract of the paper
\begin{abstract}
Using cosmological, radiation-hydrodynamic simulations targeting a rare $\approx \, 2 \times 10^{12} \, \rm M_\odot$ halo at $z \, = \, 6$, we show that the number counts and internal properties of satellite galaxies within the massive halo are sensitively regulated by a combination of local stellar radiative feedback and strong tidal forces. Radiative feedback operates before the first supernova explosions erupt and results in less tightly-bound galaxies.
Satellites are therefore more vulnerable to tidal stripping when they accrete onto the main progenitor and are tidally disrupted on a significantly shorter timescale. Consequently, the number of satellites with $M_{\rm \star} > 10^{7} \, \rm M_\odot$ within the parent system's virial radius drops by up to $60 \%$ with respect to an identical simulation performed without stellar radiative feedback. Radiative feedback also impacts the central galaxy, whose effective radius increases by a factor $\lesssim 3$ due to the presence of a more extended and diffuse stellar component.
We suggest that the number of satellites in the vicinity of massive high-redshift galaxies is an indication of the strength of stellar radiative feedback and and can be anomalously low in the extreme cosmic environments of high-redshift quasars.
\end{abstract}

% Select between one and six entries from the list of approved keywords.
% Don't make up new ones.

%%%%%%%%%%%%%%%%%%%%%%%%%%%%%%%%%%%%%%%%%%%%%%%%%%

%%%%%%%%%%%%%%%%% BODY OF PAPER %%%%%%%%%%%%%%%%%%

\section{Introduction}
\label{sec:Introduction}

Bright quasars at $z > 6$ are powered by accreting supermassive black holes with estimated masses of $\gtrsim 10^9 \, \rm M_\odot$ \citep{Fan:01, Wu:15, Banados:18}.
At $z \, = \,  6$, the Hubble time ($t_{\rm H} \, \approx \, 930 \, \rm Myr$) corresponds to $\lesssim 20$ e-folding times $t_{\rm BH} \, \approx \, 45 \left( \eta_{\rm r} / 0.1 \right) \, \rm Myr$, assuming Eddington-limited black hole accretion and a fixed radiative efficiency $\eta_{\rm r}$.
Growth to the required masses may thus proceed at the Eddington rate from the $\sim 100 \, \rm M_\odot$ remnants of Pop III stars or, alternatively, through super-Eddington accretion-limited episodes or directly from massive seed black holes with $M_{\rm BH} \, \approx \, 10^4 \-- 10^5 \, \rm M_\odot$ \citep[e.g.][]{Begelman:06}.
In all scenarios, gas inflow into the sphere of influence of the accreting black holes must be efficient, a condition which is more likely to be fulfilled if massive black holes grow at the centre of massive dark matter haloes \citep{Efstathiou:88}.

Cosmological hydrodynamic simulations following black hole accretion and active galactic nucleus (AGN) feedback have successfully reproduced the rapid assembly of $\sim 10^9 \, \rm M_\odot$ black holes by $z \, = \, 6$ \citep{Sijacki:09, DiMatteo:12}.
Using a sample of `zoom-in' cosmological simulations, \citet{Costa:14} showed that black hole growth to the required masses is efficient only in haloes with masses $M_{\rm vir} > 10^{12} \, \rm M_\odot$.
Since they trace high-$\sigma$ peaks of the cosmic density field distribution at $z \, = \, 6$ \citep{Volonteri:06}, such haloes should exhibit a statistically significant overdensity in satellite galaxy number counts when compared to lower mass systems \citep[e.g.][]{Costa:14}, though the associated variance is considerable \citep{Habouzit:19}.

This prediction has not been confirmed by observations conclusively.
Wide-field imaging campaigns have returned ambiguous results, reporting both under- and over-densities of galaxy counts in $z \gtrsim 6$ quasar fields \citep[e.g.][]{Kim:09, Balmaverde:17, Champagne:18}.
Other observations have started to probe the properties of the satellite galaxies surrounding high-redshift quasars in greater detail.
\citet{Trakhtenbrot:17} find a number of sub-mm galaxies within a projected distance $\approx 10  \-- 50\, \rm kpc$ from three out of the six $z \, \approx \, 5$ quasar fields probed by their ALMA observations, suggesting that quasar host galaxies often experience major mergers.
Similarly, based on an ALMA survey of 25 $z > 5.95$ quasars, \citet{Decarli:17} serendipitously discovered [CII] bright galaxies around four of the quasars. 
While detected satellites tend to be nearly as massive as the quasar hosts, the upcoming James Webb Space Telescope (JWST) will enable the discovery of fainter satellites and place tighter constraints on their environments.

In this paper, we present cosmological, radiation-hydrodynamic, `zoom-in' simulations of a $\approx 2.4 \times 10^{12} \, \rm M_\odot$ halo at $z \, = \, 6$ as a likely host to a bright quasar \citep[see e.g.][]{Costa:14}. In particular, we compare simulations that include or neglect stellar radiative feedback and explore the impact of stellar radiation on the demographics and properties of satellites.
We describe our simulations in Section~\ref{sec:Simulations}, present our results in Section~\ref{sec:Results}, summarise the implications of our findings in Section~\ref{sec:Implications} and our conclusions in Section~\ref{sec:Conclusions}.
A flat $\Lambda$CDM cosmology with $\Omega_{\rm m} \, = \, 0.307$, $\Omega_{\rm \Lambda} \, = \, 0.693$, $\Omega_{\rm b} \, = \, 0.049$ and $h \, = \, 0.679$ is adopted throughout this paper.

\section{ Simulations }
\label{sec:Simulations}
We perform cosmological, radiation-hydrodynamic, `zoom-in' simulations targeting a massive halo with virial mass $M_{\rm vir} \, \approx \, 2.4 \times 10^{12} \, \rm M_\odot$ and virial radius\footnote{The virial radius is defined as the radius enclosing a mean density 200 times the critical density of the Universe.} $R_{\rm vir} \, \approx \, 59.4 \, \rm kpc$, corresponding to an angular scale of $\approx 10.2''$, at $z \, = \, 6$.
This is the second most massive halo found at $z \, = \, 6$ within a cosmological volume of comoving side length $500 h^{-1} \, \rm Mpc$, as presented in \citet{Costa:18}. 
The massive galaxies evolving in such haloes are the best candidates for hosting rare supermassive black holes with masses $M_{\rm BH} \sim 10^9 \, \rm M_\odot$ at $z \, = \, 6$, as shown in \citet[e.g.][]{Costa:14}.

We employ {\sc Ramses-RT} \citep{Teyssier:02, Rosdahl:13, Rosdahl:15b} to evolve the coupled evolution of gas hydrodynamics, radiative transfer of stellar radiation and N-body dynamics of stellar populations and dark matter, modelled with particles of mass $m_{\star} \, = \, 1.3 \times 10^4 \, \rm M_\odot$ and $m_{\rm DM} \, = \, 3 \times 10^6 \, \rm M_\odot$, respectively. 
In order to increase the numerical resolution, we refine a cell if its total enclosed mass satisfies $M_{\rm DM} + \frac{\Omega_{\rm m}}{\Omega_{\rm b}} M_{\rm bar} > 8 \times m_{\rm DM}$, where $M_{\rm DM}$ and $M_{\rm bar}$ are the total dark matter and baryonic masses in the cell, respectively. The minimum cell size is $\Delta x_{\rm min} \, \approx \, 40 \, \rm pc$.
All spatial coordinates are given in physical units.

We follow non-equilibrium cooling of hydrogen and helium (coupled to the radiative fluxes present in the simulation), metal-line cooling down to $T \, = \, 10 \, \rm K$ and star formation using a Schmidt law with a variable star formation efficiency, as described in \citet{Kimm:17}.  
In order to model the ionising flux of external sources, we adopt the spatially homogeneous and time-evolving UV background of \citet{Faucher-Giguere:09}.
Supernova feedback is modelled through injection of thermal energy, if the Sedov-Taylor phase is resolved, and momentum otherwise, using the formulation of \citet{Kimm:15}.
Supernova events occur $10 \, \rm Myr$ after their parent stellar particle forms.

We focus on two simulations which differ only in whether they include or neglect stellar radiative feedback. All other parameters are kept unchanged and all other feedback processes are modelled identically. We name our simulation with radiative transfer \texttt{SN+RT} and our simulation without stellar radiation \texttt{SN}.

In \texttt{SN+RT}, we follow photo-ionisation, photo-heating and radiation pressure from stellar radiation.
Time-integration in {\sc Ramses-RT} is performed explicitly, such that the time-step is limited by the speed-of-light. In order to prevent our simulations from becoming computationally prohibitive, we adopt a reduced speed of light of $0.03 c$, which has been shown in AppendixD of \citet{Rosdahl:15} to result in well converged stellar masses, morphologies, outflow rates and ISM properties.
The emission spectra are discretised into five radiation bins as in \citet{Rosdahl:15}, i.e. infrared and optical radiation, which couple to gas solely through radiation pressure on dust, and three UV radiation groups with lower energy limits corresponding to the ionising potentials of $\rm H^+$, $\rm He^+$ and $\rm He^{++}$ \citep[as in][]{
Rosdahl:15}.
UV radiation couples to gas through radiation pressure on dust, ionising radiation pressure, photo-heating and photo-ionisation.
Radiation pressure on dust is treated both in the single- and multi-scattering regimes, for which we select specific opacities of $\kappa_{\rm ss} \, = \, 10^3 \left( Z / Z_\odot \right) \, \rm cm^{2} \, g^{-1}$, where $Z$ is the gas-phase metallicity, and $\kappa_{\rm ms} \, = \, 10 \left( Z / Z_\odot \right) \, \rm cm^{2} \, g^{-1}$, respectively.

Our aim is to quantify the differential effect of stellar radiation on the structure of a massive, high-z galaxy. AGN-driven outflows are typically far more powerful than those driven by supernovae \citep[e.g][]{Costa:14}. In addition, supernovae- and AGN-driven outflows can interact non-linearly \citep{Costa:15, Biernacki:18}, which would prevent us from isolating the impact stellar radiative feedback cleanly. We therefore exclude radiation and mechanical feedback from AGN and black hole growth from our simulations.

Haloes and galaxies are identified using {\sc AdaptaHop} \citep{Aubert:04, Tweed:09} in the most massive submaxima (MSM) mode.
We require a minimum of 20 particles per halo.
Haloes are selected from matter overdensities higher than $\rho_{\rm TH} \, = \, 80 \rho_{\rm mean}$, where $\rho_{\rm mean}$ is the mean density of the Universe.
We choose parameters $N_{\rm SPH} \, = \, 32$, which gives the number of nearest-neighbours used to smooth the density field around each dark matter particle, $N_{\rm HOP} \, = \, 16$, as the number of particle neighbours used to determine the density gradient around each particle when assigning it to its local patch, and $f_{\rm Poisson} \, = \, 4$, which ensures that only clumps identified at $4\sigma$ significance are retained in our catalogues \citep[see AppendixB in][for a description of all parameters]{Aubert:04}. 
We follow two approaches in identifying galaxies: (i) we assign galaxies to haloes by adding up the masses of all stellar particles found within $30 \%$ of a given halo's virial radius \citep[as in][]{Rosdahl:18} and, as an alternative, (ii) we explore identifying galaxies from the stellar particle field directly using {\sc AdaptaHop} with $\rho_{\rm TH} \, = \, 200 \rho_{\rm mean}$, $N_{\rm SPH} \, = \, 20$, $N_{\rm HOP} \, = \, 20$ and $f_{\rm Poisson} \, = \, 4$.

\section{ Results }
\label{sec:Results}
\begin{figure}
	\includegraphics[width=0.5\textwidth]{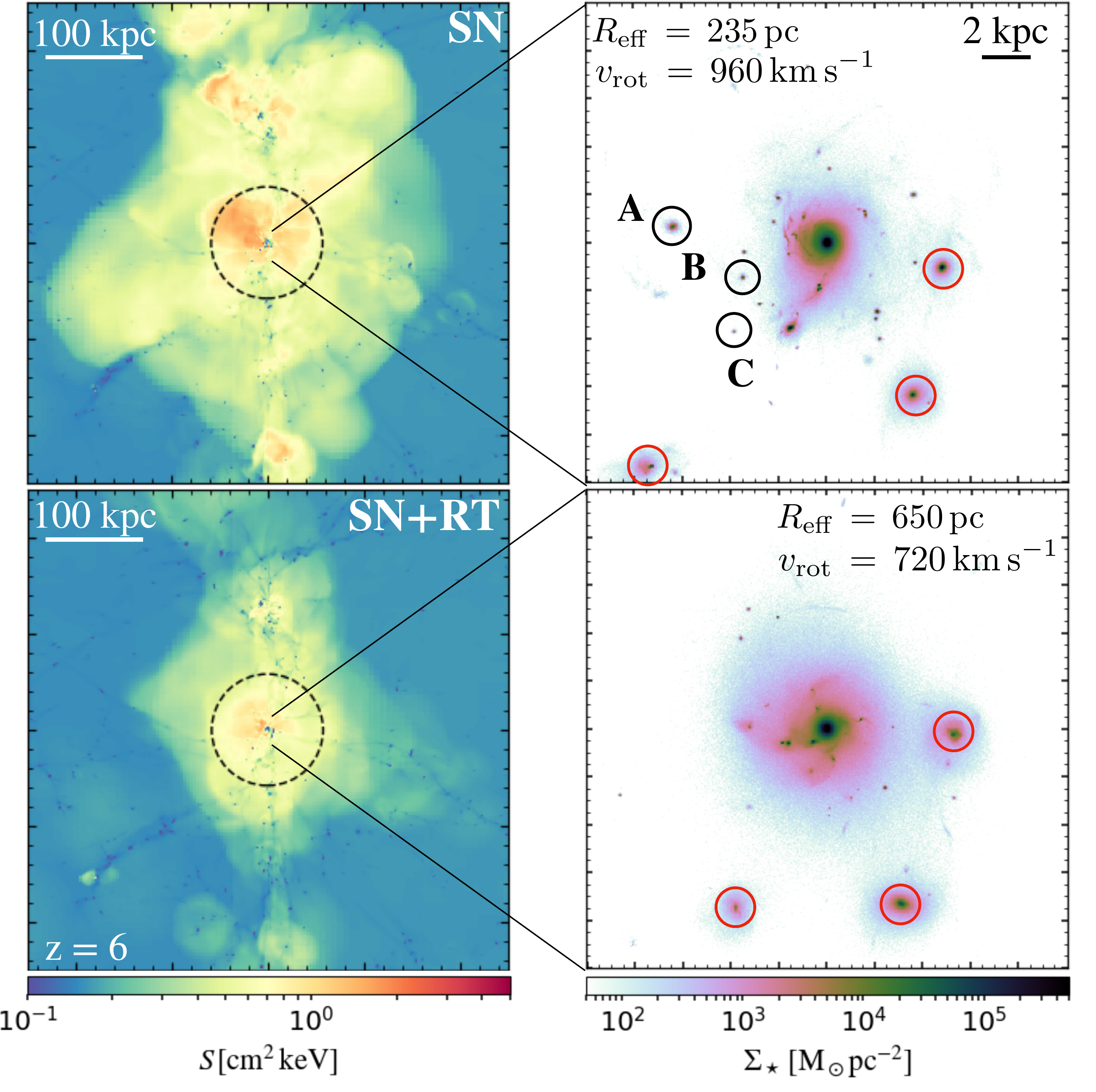}
	\caption{Mass-weighted entropy within a cubic volume of side length $500 \rm kpc$ centred on the most massive galaxy at $z \, = \, 6$ in the simulation without stellar radiative feedback (top, left panel) and with stellar radiation (bottom, left panel). The black circle marks the halo's virial radius. By suppressing star formation, stellar radiation leads to weaker supernova-driven outflows, explaining the lower entropies seen in the simulation with radiative feedback. In the right-hand panels, we show the stellar surface density in the central few kpc of the massive galaxy. In the simulation without stellar radiation, the targeted galaxy is more compact, spins faster and is significantly clumpier. In the simulation with radiative feedback, the massive galaxy is more spatially extended and has a significantly more diffuse and smoother stellar component. The clumps highlighted with black circles are example remnants of accreted satellites, which are more thoroughly destroyed in the simulation with stellar radiation (see text). The red circles show examples of satellite galaxies that exist in both simulations.}\
	\label{fig_main}
\end{figure}

In Fig.~\ref{fig_main}, we show the entropy distribution of gas within a cube of side length $500 \, \rm kpc$ centred on the target galaxy at $z \, = \, 6$.
The top panel, which shows results for \texttt{SN}, shows a prominent bubble of high entropy gas that extends out to $\approx 3 \--  5 R_{\rm vir}$ ($\approx 180 \-- 300 \, \rm kpc$), where $R_{\rm vir}$ is indicated with a circle. 
The bubble, which is composed of hot gas heated by supernova-driven blasts and accretion shocks, encircles the targeted galaxy as well as other massive satellites in its vicinity.

If radiative feedback is included, both the spatial scale of the high entropy region and the typical entropy within the bubble lessen substantially, as shown in the bottom panel of Fig.~\ref{fig_main}; the importance of shocks diminishes in the presence of stellar radiation.
High entropy gas becomes rarer, because supernova-driven outflows become weaker in the presence of stellar radiation. On the one hand, this is because the total stellar mass in \texttt{SN+RT} is at most times lower by $20 \-- 40 \%$, such that there are fewer supernova events overall \citep{Rosdahl:15, Agertz:19} and, on the other hand, because supernova explosions become less spatially and temporally correlated \citep[e.g.][]{Kimm:18}.

The stronger supernova-driven winds generated in the absence of radiative feedback occur in response to the more efficient gas collapse that takes place in \texttt{SN}, where the absence of processes regulating gas accretion onto star-forming sites before supernova feedback operates \citep[see e.g.][]{Peters:17} results in generally denser stellar structures. 
At fixed halo mass, we find moderate but systematic enhancements in stellar mass already at $z \approx 8$; the mean stellar mass of all galaxies in the high-resolution volume is higher by $0.1 \-- 0.2 \, \rm dex$ in \texttt{SN} than in \texttt{SN+RT} across the full halo mass range. More strikingly, galaxies are typically more tightly-bound in \texttt{SN} than in \texttt{SN+RT} at these early times; we find a mean peak stellar circular velocity of $v_{\rm pk} \approx \, 65 \, \rm km s^{-1}$ in \texttt{SN}, compared to $v_{\rm pk} \approx \, 35 \, \rm km s^{-1}$ in \texttt{SN+RT} in galaxies hosted by haloes with $M_{\rm vir} > 10^{10} \, \rm M_\odot$.

Such differences in galactic internal structure persist at lower redshift and are pronounced also for the most massive galaxy.
At $z \, \approx \, 6$, we find that the star formation rate in \texttt{SN+RT} starts exceeding that of \texttt{SN}, as the gas which fails to form stars in progenitor galaxies due to radiative feedback at higher redshift undergoes star formation in the massive system instead. 
Accordingly, the stellar mass of the massive galaxy at $z \, = \, 6$, estimated by adding up the masses of all stellar particles within $30\%$ of the host's virial radius, is $\approx 1.7 \times 10^{11} \, \rm M_\odot$ in \texttt{SN} and $\approx 1.5 \times 10^{11} \, \rm M_\odot$ in \texttt{SN+RT}. The difference in the target galaxy's stellar mass is more significant at higher redshift, e.g. $\approx 1.1 \times 10^{11} \, \rm M_\odot$ in \texttt{SN} and $\approx 7.5 \times 10^{10} \, \rm M_\odot$ in \texttt{SN+RT} at $z \, = \, 6.5$.
At no point in the simulation, however, does the difference in the host stellar mass exceed $40 \%$ between the two simulations. Accordingly, the stellar-to-halo mass ratio for the targeted galaxy is $M_{\rm \star} / M_{\rm vir} \approx 0.07$ and $M_{\rm \star} / M_{\rm vir} \approx 0.06$ in \texttt{SN} and \texttt{SN+RT}, respectively. Both values are in reasonable agreement with the abundance matching expectations of \citet{Moster:18} who predict $M_{\rm \star} / M_{\rm vir} \approx 0.04$ at $z \, = \, 6$, somewhat higher than those of \citet{Behroozi:13} ($M_{\rm \star} / M_{\rm vir} \approx 0.01$) but in very close agreement with the recent simulations of a similarly massive galaxy at  $z \, = \, 7$ of \citet{Lupi:19}, where $M_{\rm \star} / M_{\rm vir} \approx 0.06$.

The right-hand panels of Fig.~\ref{fig_main} display the stellar surface density fields around the targeted system.
With a stellar half-mass radius of $R_{\rm eff} \, = \, 235 \, \rm pc$, the stellar component is more concentrated in \texttt{SN} than in \texttt{SN+RT}, where $R_{\rm eff} \, = \, 650 \, \rm pc$, a size difference of a factor $\lesssim 3$.
In addition, the mean stellar rotational velocity around the galaxy's angular momentum vector, averaged over the redshift range $6 < z < 6.25$  and evaluated within $R_{\rm eff}$ equals $v_{\rm rot} \, \approx \, 960 \, \rm km \, s^{-1}$ in \texttt{SN} and $v_{\rm rot} \, \approx \,720 \, \rm km \, s^{-1}$ in \texttt{SN+RT}.
That other systems are also more tightly-bound in \texttt{SN} can be seen directly from the stellar surface density maps shown in Fig.~\ref{fig_main}, where the central surface densities of the satellites are typically higher in \texttt{SN} than in \texttt{SN+RT}.
There is also a clear excess of compact ``clumps" in the simulation without radiative feedback, which makes the galaxy of \texttt{SN} appear to be richer in structure than in \texttt{SN+RT}, where the stellar component looks decidedly smoother.
Stellar radiation results in less tightly-bound, somewhat less massive galaxies \citep[in agreement with][]{Rosdahl:15, Kimm:18, Hopkins:18} as well as in more spatially extended, ``puffed-up'' systems with less structure.

The clumps appearing in the stellar surface density map on the top, right-hand panel of Fig.~\ref{fig_main} have typical masses of $10^7 \-- 10^9 \, \rm M_\odot$; for instance clumps A, B and C marked in Fig.~\ref{fig_main} have approximate masses of $9 \times 10^8 \, \rm M_\odot$, $2 \times 10^8 \, \rm M_\odot$ and $8 \times 10^7 \, \rm M_\odot$, respectively. 
While the additional stellar systems, including clumps A, B and C may, at first glance, resemble the stellar clumps that form in-situ within simulated high-redshift galaxies \citep[e.g.][]{Mandelker:17}, we find, by tracing their constituent stellar particles back in time, that many of them form at $z \approx 7 \-- 9$ in separate dark matter haloes $10 \-- 100 \, \rm kpc$ away from the main progenitor.
For instance, clumps A, B and C in Fig.~\ref{fig_main} form half of their stellar mass at $z \, \approx \, 8$ within dark matter haloes with $M_{\rm vir} \, \sim 10^{10} \, \rm M_\odot$. There are no evident counterparts for these clumps for \texttt{SN+RT}, even though this simulation also contains some (but markedly fewer) clumps and some satellites that also exist in \texttt{SN} (red circles).

\begin{figure}
	\includegraphics[width=0.475\textwidth]{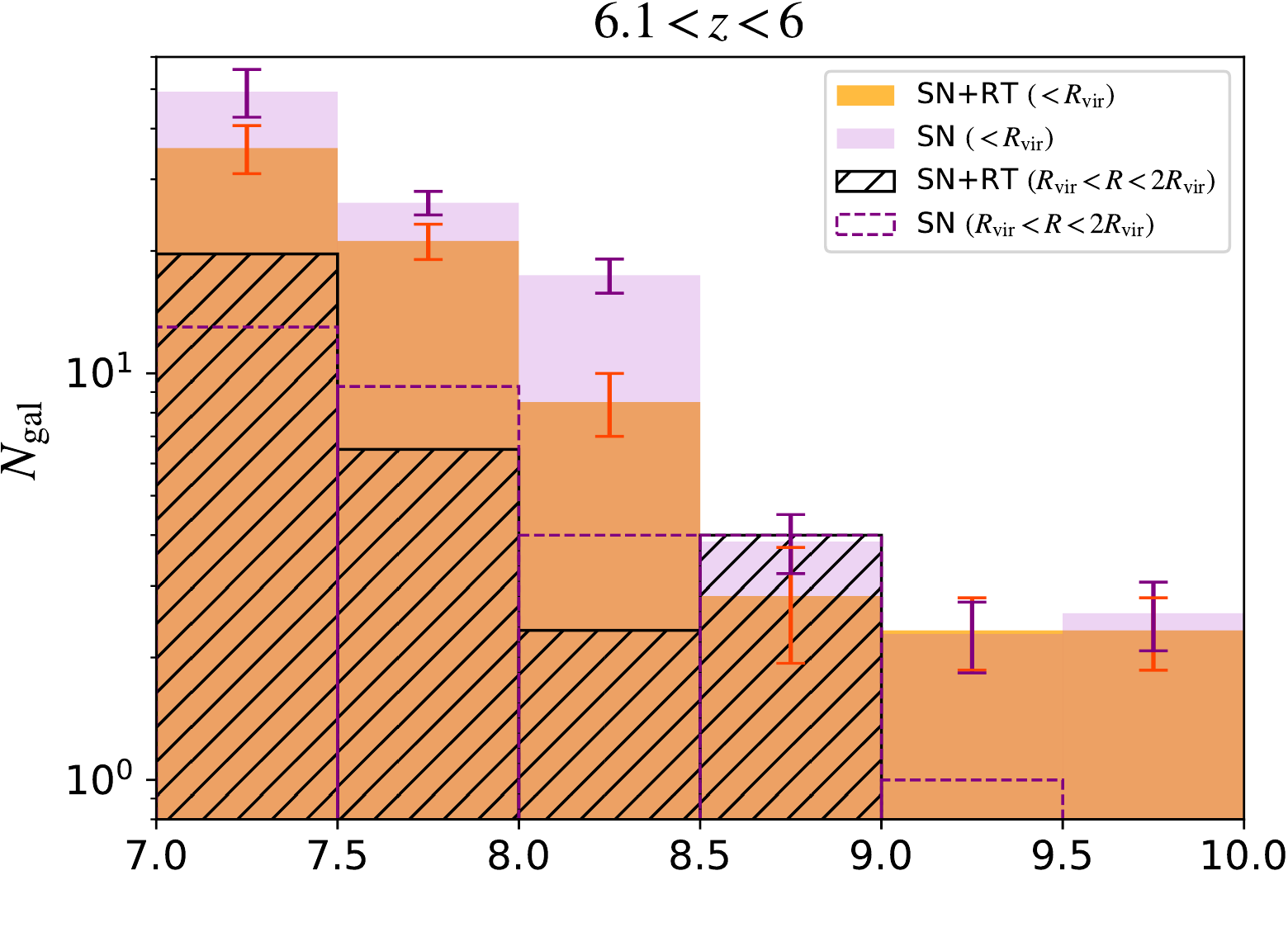}
	\includegraphics[width=0.475\textwidth]{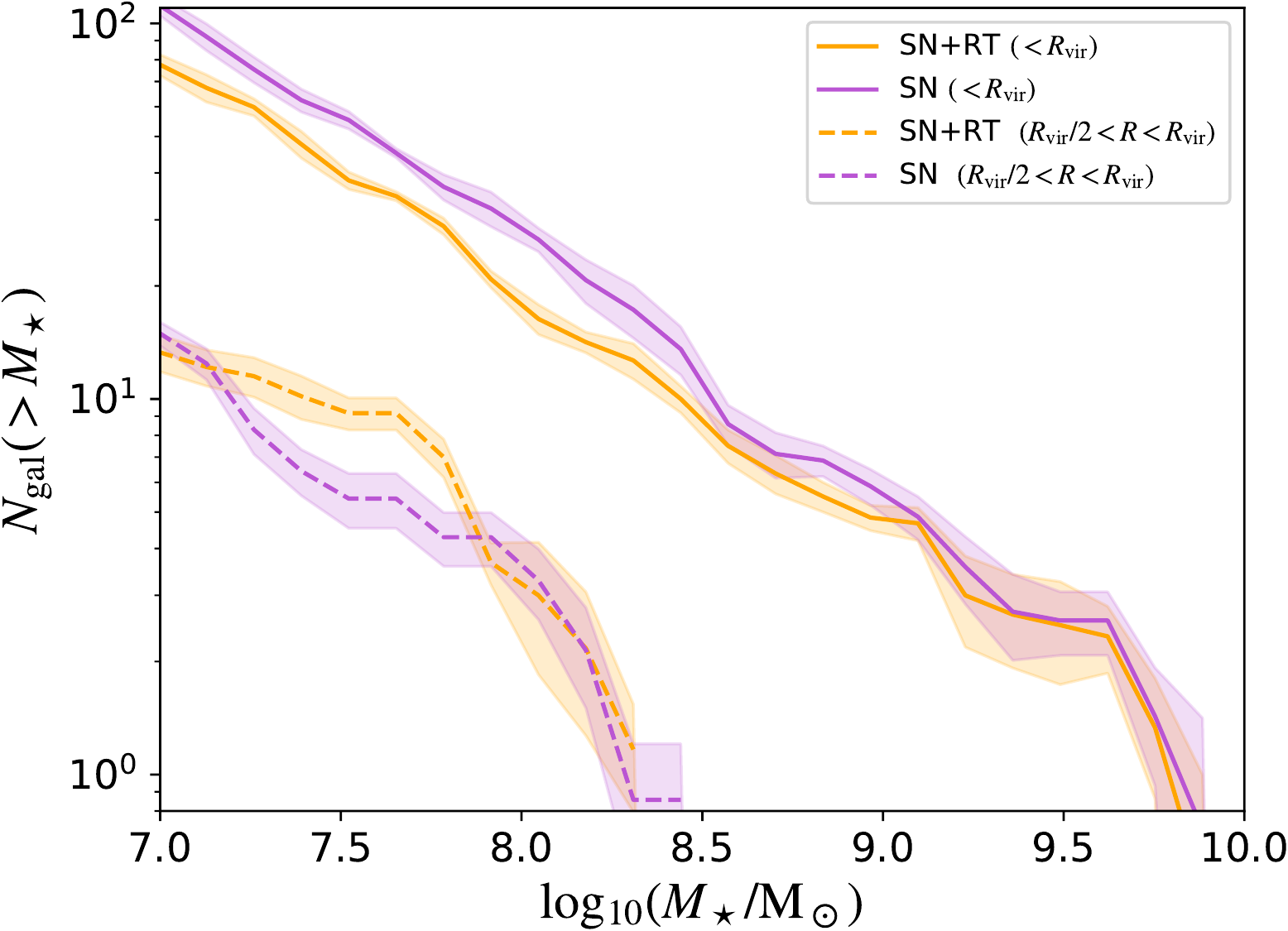}
	\caption{Top: Average number of satellites per logarithmic stellar mass within $R_{\rm vir}$ (filled histograms) and in the radius range $R_{\rm vir} < R < 2R_{\rm vir}$ (open and hatched histograms), as obtained by averaging over all simulations snapshots between $z \, = 6.1$ and $z \, = \, 6$. Violet histograms show the satellite number count in the simulation without stellar radiative feedback, while the orange and hatched histograms give the result for the simulation following stellar radiation self-consistently. Bottom: Mean cumulative number of galaxies above any given stellar mass at $z \, = \, 6$ within $R_{\rm vir}$ (solid curves) and within $R_{\rm vir}/2 < R < R_{\rm vir}$ (dashed curves). Stellar radiation results in a significantly smaller satellite galaxy population within the virial radius. The mean number of systems with $M_{\rm \star} > 10^7 (10^8) \, \rm M_\odot$ is $102 \pm 8$ ($26 \pm 2$) in \texttt{SN} and $73 \pm 5$ ($16 \pm 2$) in \texttt{SN+RT}; the number of galaxies within the halo drops by up to $40\%$ with the addition of stellar radiation. Error bars and shaded regions denote $1 \sigma$ intervals.}
	\label{fig_nsat}
\end{figure}

\begin{figure*}
	\includegraphics[width=0.98\textwidth]{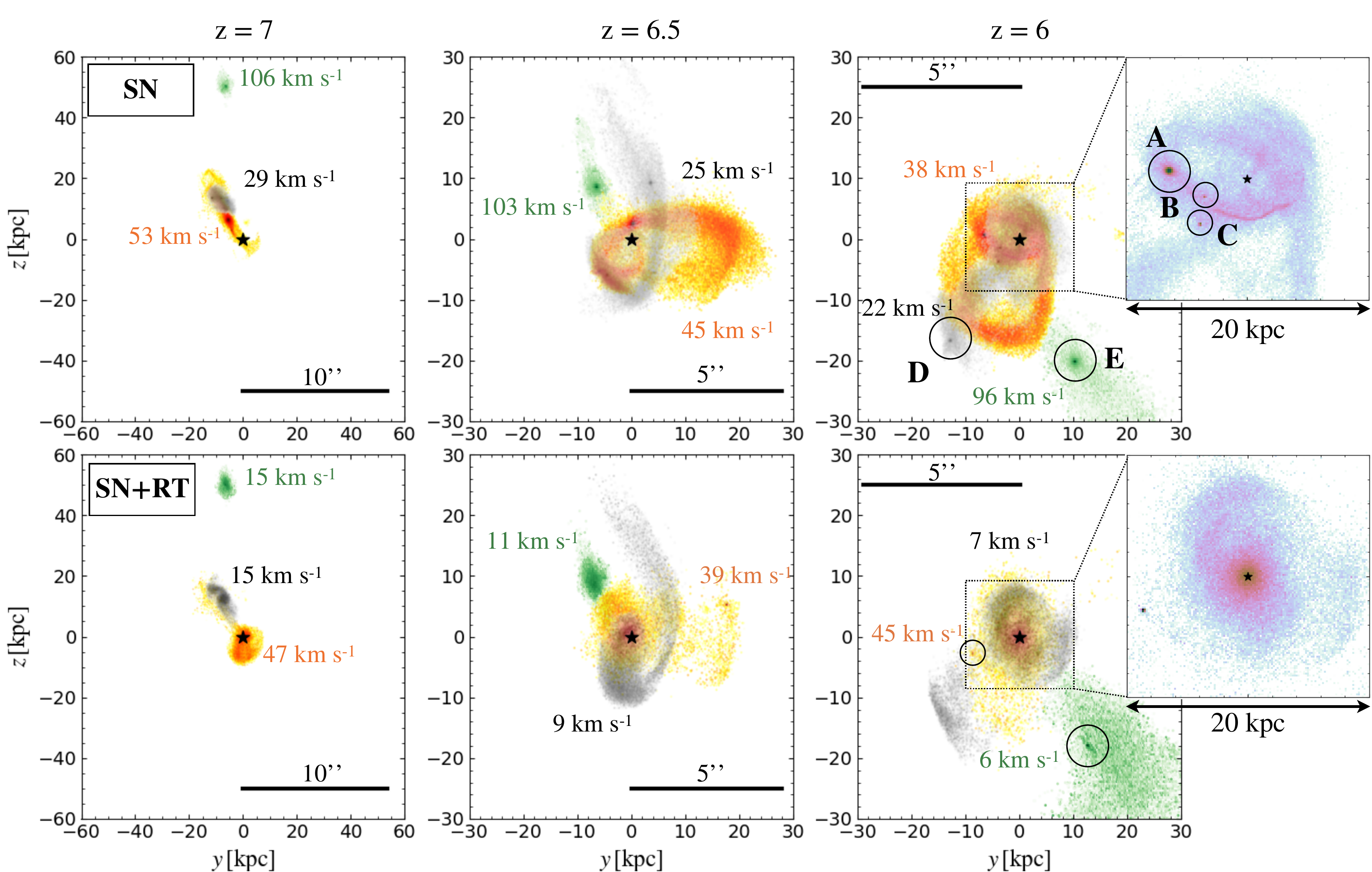}
	\caption{We select three systems with $M_{\rm vir} \sim 10^{10} \, \rm M_\odot$ at $z \, = \, 8$, matching them between the simulations without- and with radiative feedback. We track the stellar particles within $30\% \, R_{\rm vir}$ forward in time and show the resulting spatial distribution at $z\, = \,  7$ (first column), $z\, = \,  6.5$ (second column) and $z\, = \,  6$ (third column) in \texttt{SN} (top row) and \texttt{SN+RT} (bottom row). We provide the peak stellar circular velocity next to each system, taking the location of the system's centre as the position of the most tightly-bound stellar particle. The galaxies undergo significant tidal interactions with the massive central, exhibiting prominent tidal tails. Crucially, the less tightly-bound galaxies in \texttt{SN+RT} are more easily tidally disrupted, often dispersing entirely. In \texttt{SN}, all selected galaxies survive in the form of compact cores, including clumps A, B and C identified in Fig.~\ref{fig_main}. In the rightmost panels, we show the spatial distribution of all tracked stellar particles at $z \, = \, 6$. Many of the stellar clumps orbiting around the massive galaxy are the tightly-bound cores of accreted satellites which, by virtue of their lower binding energy, are less long-lived in the presence of stellar radiation.}\
	\label{fig_disruption}
\end{figure*}

We are thus led to investigate why there may be hidden satellites in \texttt{SN+RT}.
We first investigate the satellite population, identified through the method (i) described in the last paragraph of Section~\ref{sec:Simulations}, at $z \, = \, 6$ in both simulations. Remarkably, we find 6 satellites with $M_{\rm \star} > 10^8 \, \rm M_\odot$ within $R_{\rm vir}$ in \texttt{SN}, compared to $2$ in \texttt{SN+RT}. The discrepancy between the number counts of massive satellites is significantly weaker if we instead select systems outside $R_{\rm vir}$; there are 12 galaxies with $M_{\rm \star} > 10^8 \, \rm M_\odot$ between $R_{\rm vir}$ and $3 \times R_{\rm vir}$ in \texttt{SN} and 11 in \texttt{SN+RT}. If we select systems from within the whole high-resolution region, excluding those within $R_{\rm vir}$, the discrepancy in number counts is $\lesssim 5 \%$. The difference in the number count of massive systems is amplified close to the quasar host galaxy.

Accreted satellites, however, often lose their dark matter haloes due to strong tidal interactions with the massive central, such that method (i) does not pick out many of the clumps seen in in Fig.~\ref{fig_main}.
In the top panel of Fig.~\ref{fig_nsat}, we plot the number of galaxy satellites, as identified directly using {\sc AdaptaHop}, i.e. the method (ii) described in Section~\ref{sec:Simulations}, as a function of stellar mass.
The values shown in the top panel of Fig.~\ref{fig_nsat} correspond to the mean number, as obtained by averaging over all the snapshots in the redshift range $6.1 < z < 6$.
Excluding stellar radiation leads to a significant excess in the number of galaxies within $R_{\rm vir}$ in almost every stellar mass bin, but in particular for $M_{\rm \star} < 10^9 \, \rm M_\odot$. 
The decrease in galaxy number counts is also clear in the bottom panel of Fig.~\ref{fig_nsat}, where cumulative galaxy counts are shown for $z \, = \ 6$ within two different spatial scales.
The excess is of $\approx 30$ massive clumps with $M_{\rm \star} > 10^7 \, \rm M_\odot$ or $\approx 10$ with $M_{\rm \star} > 10^8 \, \rm M_\odot$, within the virial radius. 
Both for $M_{\rm \star} > 10^7 \, \rm M_\odot$ or $M_{\rm \star} > 10^8 \, \rm M_\odot$, the inclusion of stellar radiation thus leads to a $\approx 40 \% \-- 60\%$ reduction in the number of satellites and stellar clumps within the virial radius.
Notably, the distributions of galaxy counts as a function of stellar mass are similar between \texttt{SN} and \texttt{SN+RT} (open and hatched histograms in the top panel and dashed curves in the bottom panel of Fig.~\ref{fig_nsat}) in the outskirts of the halo and beyond the virial radius.

In order to gain insight into why there is a loss of structure when we include stellar radiation into our simulations, we select three dark matter haloes with $M_{\rm vir} \sim 10^{10} \, \rm M_\odot$ at $z \, = \, 8$, that are known to end up within the most massive halo at $z \, = \, 6$, and extract all the stellar particles residing within $30 \% \, R_{\rm vir}$.
By matching the IDs of dark matter particles in \texttt{SN} and \texttt{SN+RT}, we ensure we identify the same halo in both simulations.
Fig.~\ref{fig_disruption} shows the spatial distribution of the selected stellar particles at three different redshifts (shown in different columns) in \texttt{SN} (top row) and in \texttt{SN+RT} (bottom row).
While the spatial configuration of the stellar component is similar between both simulations at $z \, \gtrsim \, 7$,  their evolution differs significantly at lower redshift. 
Prominent tails of stripped material are visible both in \texttt{SN} and \texttt{SN+RT}, indicating that the selected systems are strongly tidally disrupted as they approach the central galaxy.
Crucially, various tightly-bound stellar cores survive intact in \texttt{SN}, while the selected galaxies dissipate almost entirely in \texttt{SN+RT}, forming a diffuse stellar envelope around the central galaxy.
The clearest example is the galaxy shown in green in Fig.~\ref{fig_disruption}. As it passes through the central galaxy, this system survives relatively undisturbed in \texttt{SN} (clump E), but is completely shattered in \texttt{SN+RT}.
Similarly, the galaxy shown in grey is completely shredded in \texttt{SN+RT}, while it evolves into a system of multiple tightly-bound cores in \texttt{SN} (clumps B, C also shown in Fig.~\ref{fig_main}, and D).
Finally, even if the system shown in orange survives as a core in both simulations, this has a mass of $M_{\rm \star} \, \approx \, 4 \times 10^7 \, \rm M_\odot$ in \texttt{SN+RT}, while it has $M_{\rm \star} \, \approx \, 9 \times 10^8 \, \rm M_\odot$ in \texttt{SN}.
Many of the ``stellar clumps'' seen in Fig.~\ref{fig_main} are therefore the remnant cores of accreted satellite galaxies which are destroyed in \texttt{SN+RT}, but survive in \texttt{SN}. It is clear that the difference in  satellite counts highlighted in Fig.~\ref{fig_nsat} is caused by the more efficient tidal disruption in \texttt{SN+RT}.

\section{Implications}
\label{sec:Implications}
Stellar radiation reduces the stellar masses of the galaxy population that emerges in our simulations by a factor $\lesssim 2$ \citep[in agreement with][]{Rosdahl:15, Kimm:18, Hopkins:18, Kannan:18}.
In addition, by acting on surrounding gas immediately after the formation of stellar populations, stellar radiation counters gas collapse in galaxies residing in haloes with $M_{\rm vir} \lesssim 10^{10} \, \rm M_\odot$, such that they become more diffuse and less tightly-bound.

At $z > 6$, systems with stellar masses of $M_\star \sim 10^8 \-- 10^9 \, \rm M_\odot$ typically represent central galaxies. In our simulations, they instead consist of the progenitors of a massive galaxy capable of harbouring a quasar, with many such systems orbiting within the massive halo as satellites.
The deep gravitational potential well of the massive galaxy, with high peak stellar circular velocities of $v_{\rm pk} > 700 \, \rm km \, s^{-1}$ rapidly disrupts all but the most tightly-bound of the accreted systems.
We have shown that this process is more efficient if early feedback operates.

Our findings have potentially far-reaching implications for future observational missions aiming at probing the properties and number counts of satellites within close proximity of high-redshift massive galaxies and quasars. These may be fewer in number than na\"ively expected, a result which could be misinterpreted as an indication that the mass of the dark matter haloes hosting $z > 6$ quasars must be lower than typically assumed.
The preprocessing of low-mass galaxies at high-redshift by stellar radiation may also reduce the number of satellites around low-redshift galaxies, though we expect the shredding of these satellites to become less efficient than seen for the massive $z \, = \, 6$ haloes presented here, due to the weaker tidal fields. It will be important to test the mechanism outlined in this paper in lower mass haloes and at lower redshift.

Due to efficient tidal stripping, a significant number of the neighbouring systems of $z > 6$ massive galaxies and quasars should also appear irregular and, in the likely case that only the tightly-bound cores are detected, unusually compact.
Based on our results, we also expect quasar host galaxies to be surrounded by a rich web of stellar streams composed of the tidally stripped stellar components of accreted dwarf galaxies and to contain a diffuse stellar component out to radii of $\approx 10 \, \rm kpc$, just under $2''$. 

The different satellite destruction timescales also carry implications for the morphology of the massive galaxy. Early radiative feedback leads to satellites which are destroyed on a timescale shorter than the dynamical friction timescale and therefore add to the diffuse stellar halo that surrounds the massive galaxy already at $z \, = \, 6$. The more tightly-bound satellites that would form otherwise are more likely to sink into the centre of the galaxy by dynamical friction, contributing to the growth of the central bulge instead.

Future work should therefore explore the evolution of massive galaxies such as that explored here down to $z \, = \, 0$. According to the few cosmological, radiation-hydrodynamic simulations performed down to $z \, = \, 0$, stellar radiation operates efficiently in dwarfs with $M_{\rm vir} \lesssim 10^{10} \, \rm M_\odot$ during re-ionisation \citep{Agertz:19, Katz:19}, but introduces only modest corrections to bulk quantities such as stellar mass in more massive systems with $M_{\rm vir} \sim 10^{12} \, \rm M_\odot$ as well as moderately flatter stellar density profiles \citep[e.g.][]{Hopkins:18} by $z \, = \, 0$. \citet{Rosdahl:15}, \citet{Kimm:18} and \citet{Hopkins:18} highlight how stellar radiation, however, results in a smaller number of tightly-bound star clusters, such that the clearest signature of stellar radiation may be in the morphology of the galaxy and the properties of its star forming regions. If we extrapolate these findings to the galaxy studied in this paper, we may expect stellar radiation to contribute negligibly to the final stellar mass of the galaxy, which is more likely set by AGN feedback, but to significantly reduce the central stellar density and the number of star clusters and satellites in its vicinity.

While we have accounted for stellar radiation in our simulations, we have neglected the radiation field of the quasar itself, which may have a strong impact on the ability of gas to accrete onto dwarf galaxies, leading to potentially lower mass and less tightly-bound progenitors.
In previous simulations \citep[see][]{Costa:18} we followed quasar radiation in the same halo at $z \, < \, 6.5$ and found no difference in the satellite demographics. However, if quasar radiation is `switched-on' at an earlier time, it remains possible for the process outlined here to be amplified.

In addition, since stellar radiation regulates star formation to some extent, we may na\"ively expect that it could, in turn, suppress black hole accretion, particularly in the lower mass progenitors at very high-z. However, since stellar radiation (i) stabilises larger masses of gas against collapse and (ii) results in weaker large-scale outflows, we find that the massive galaxy at $z \, = \, 6$, as well as its progenitors contain \emph{larger} central gas reservoirs in \texttt{SN+RT} out to $z \gtrsim 10$ (see Table~\ref{table:free_parameters}), a time before most black hole growth is likely to occur \citep[see][]{Costa:14}. This counter-intuitive results echoes one of the findings of \citet{Costa:14}, who showed that more highly suppressed star formation due to stronger stellar feedback leads, counter-intuitively, to a higher black hole mass. The higher abundance of cold gas within the central regions of the targeted, massive galaxy and its progenitors, in fact, suggests that stellar radiation may aid in growing supermassive black holes by $z \, = \, 6$, an exciting possibility that deserves to be explored in a future study.

\begin{table}
\begin{center}
\begin{tabular}{l*{2}{c}c}
\hline
Simulation  & $z = 10$ & $z = 8$ & $z = 6$ \\
\hline
\texttt{SN}        & $7.4 \times 10^8 \rm \, M_\odot$ & $2.4 \times 10^9 \rm \, M_\odot$  & $1.1 \times 10^{10} \rm \, M_\odot$  \\
\texttt{SN+RT} & $1.5 \times 10^9 \rm \, M_\odot$ & $5.9 \times 10^9 \rm \, M_\odot$  & $3.7 \times 10^{10} \rm \, M_\odot$  \\
\hline
\end{tabular}
\caption{The total gas mass in the central kpc of the most massive galaxy at different redshifts in \texttt{SN} (top) and \texttt{SN+RT} (bottom). There is a systematic excess of gas in the simulation with stellar radiative feedback, where both star formation and outflows are suppressed simultaneously. We speculate that the excess gas in the galactic nucleus can fuel more intense black hole growth.}
\label{table:free_parameters}
\end{center}
\end{table}

Besides investigating the impact of quasar radiation on the satellite population, it will be important to simulate a larger number of massive $z \, = \, 6$ haloes to quantify the impact of cosmic variance on our findings.
Increasing resolution will, in turn, allow for yet more detailed studies of the internal structure and kinematics of satellite galaxies.

\section{Summary}
\label{sec:Conclusions}

Using cosmological, radiation-hydrodynamic simulations targeting a rare $M_{\rm vir} \, \approx \, 2 \times 10^{12} \, \rm M_\odot$ halo at $z \, = \, 6$, we have shown that stellar radiative feedback results in less tightly-bound central and satellite galaxies. As opposed to supernovae, which take $\sim 10 \, \rm Myr$ to erupt, stellar radiation counters gas accretion immediately after stars form. Stellar radiative feedback is gentle, but has surprising effects: it results in weaker supernova feedback and in less structure.
While it does not strongly alter the demographics of galaxies residing far from the massive targeted galaxy, stellar radiative feedback is ultimately responsible for a smaller satellite population within the latter's virial radius.
Before they sink into the galactic nucleus on a dynamical friction timescale, the less dense satellites are efficiently tidally disrupted and incorporated into the diffuse stellar halo that envelopes the quasar host galaxy.
We suggest that this anomaly is a unique imprint of the extreme environments of high-redshift galaxies and bright quasars at $z \, = \, 6$ and should be sensitive to the host halo mass.
Its magnitude should, in turn, be a measure of the importance of early feedback in galaxy formation.
Crucially, our findings suggest that future missions, performed with instruments such as JWST, may unveil a surprisingly low number of galactic satellites with $M_{\rm \star} > 10^7 \, \rm M_\odot$ around $z > 6$ quasars, which could be erroneously attributed to a lower parent halo mass.

\section*{Acknowledgements}
The authors thank the anonymous referee for a useful and constructive report.  
TC is grateful to Benny Trakhtenbrot for thoroughly reading the manuscript and for providing many insightful comments that greatly improved its clarity.
TC further acknowledges Martin Haehnelt, Thorsten Naab, R\"udiger Pakmor, Debora Sijacki, Volker Springel and Freeke van de Voort for helpful comments and discussions.
JR acknowledges support from the ORAGE project from the Agence Nationale de la Recherche under grant ANR-14-CE33-0016-03.
TK was supported in part by the National Research Foundation of Korea (No. 2017R1A5A1070354 and No. 2018036146) and in part by the Yonsei University Future-leading Research Initiative (RMS2-2018-22-0183).
This work was partially carried out on the Dutch national e-infrastructure with the support of SURF cooperative. We acknowledge that the results of this research have been achieved using the DECI resource Eagle, at PSNC in Poland with support from the PRACE aisbl.
%%%%%%%%%%%%%%%%%%%%%%%%%%%%%%%%%%%%%%%%%%%%%%%%%%

%%%%%%%%%%%%%%%%%%%% REFERENCES %%%%%%%%%%%%%%%%%%

% The best way to enter references is to use BibTeX:

\bibliographystyle{mnras}
\bibliography{lit} % if your bibtex file is called example.bib

\begin{thebibliography}{}
\makeatletter
\relax
\def\mn@urlcharsother{\let\do\@makeother \do\$\do\&\do\#\do\^\do\_\do\%\do\~}
\def\mn@doi{\begingroup\mn@urlcharsother \@ifnextchar [ {\mn@doi@}
  {\mn@doi@[]}}
\def\mn@doi@[#1]#2{\def\@tempa{#1}\ifx\@tempa\@empty \href
  {http://dx.doi.org/#2} {doi:#2}\else \href {http://dx.doi.org/#2} {#1}\fi
  \endgroup}
\def\mn@eprint#1#2{\mn@eprint@#1:#2::\@nil}
\def\mn@eprint@arXiv#1{\href {http://arxiv.org/abs/#1} {{\tt arXiv:#1}}}
\def\mn@eprint@dblp#1{\href {http://dblp.uni-trier.de/rec/bibtex/#1.xml}
  {dblp:#1}}
\def\mn@eprint@#1:#2:#3:#4\@nil{\def\@tempa {#1}\def\@tempb {#2}\def\@tempc
  {#3}\ifx \@tempc \@empty \let \@tempc \@tempb \let \@tempb \@tempa \fi \ifx
  \@tempb \@empty \def\@tempb {arXiv}\fi \@ifundefined
  {mn@eprint@\@tempb}{\@tempb:\@tempc}{\expandafter \expandafter \csname
  mn@eprint@\@tempb\endcsname \expandafter{\@tempc}}}

\bibitem[\protect\citeauthoryear{{Agertz} et~al.,}{{Agertz}
  et~al.}{2019}]{Agertz:19}
{Agertz} O.,  et~al., 2019, \mnras \, (submitted), \href
  {https://ui.adsabs.harvard.edu/abs/2019arXiv190402723A} {p. arXiv:1904.02723}

\bibitem[\protect\citeauthoryear{{Aubert}, {Pichon}  \& {Colombi}}{{Aubert}
  et~al.}{2004}]{Aubert:04}
{Aubert} D.,  {Pichon} C.,   {Colombi} S.,  2004, \mn@doi [\mnras]
  {10.1111/j.1365-2966.2004.07883.x}, \href
  {https://ui.adsabs.harvard.edu/abs/2004MNRAS.352..376A} {352, 376}

\bibitem[\protect\citeauthoryear{{Ba{\~n}ados} et~al.,}{{Ba{\~n}ados}
  et~al.}{2018}]{Banados:18}
{Ba{\~n}ados} E.,  et~al., 2018, \mn@doi [\nat] {10.1038/nature25180}, \href
  {https://ui.adsabs.harvard.edu/\#abs/2018Natur.553..473B} {553, 473}

\bibitem[\protect\citeauthoryear{{Balmaverde} et~al.,}{{Balmaverde}
  et~al.}{2017}]{Balmaverde:17}
{Balmaverde} B.,  et~al., 2017, \mn@doi [\aap] {10.1051/0004-6361/201730683},
  \href {http://adsabs.harvard.edu/abs/2017A%26A...606A..23B} {606, A23}

\bibitem[\protect\citeauthoryear{{Begelman}, {Volonteri}  \& {Rees}}{{Begelman}
  et~al.}{2006}]{Begelman:06}
{Begelman} M.~C.,  {Volonteri} M.,   {Rees} M.~J.,  2006, \mn@doi [\mnras]
  {10.1111/j.1365-2966.2006.10467.x}, \href
  {http://adsabs.harvard.edu/abs/2006MNRAS.370..289B} {370, 289}

\bibitem[\protect\citeauthoryear{{Behroozi}, {Wechsler}  \&
  {Conroy}}{{Behroozi} et~al.}{2013}]{Behroozi:13}
{Behroozi} P.~S.,  {Wechsler} R.~H.,   {Conroy} C.,  2013, \mn@doi [\apj]
  {10.1088/0004-637X/770/1/57}, \href
  {https://ui.adsabs.harvard.edu/abs/2013ApJ...770...57B} {770, 57}

\bibitem[\protect\citeauthoryear{{Biernacki} \& {Teyssier}}{{Biernacki} \&
  {Teyssier}}{2018}]{Biernacki:18}
{Biernacki} P.,  {Teyssier} R.,  2018, \mn@doi [\mnras] {10.1093/mnras/sty216},
  \href {https://ui.adsabs.harvard.edu/abs/2018MNRAS.475.5688B} {475, 5688}

\bibitem[\protect\citeauthoryear{{Champagne} et~al.,}{{Champagne}
  et~al.}{2018}]{Champagne:18}
{Champagne} J.~B.,  et~al., 2018, \mn@doi [\apj] {10.3847/1538-4357/aae396},
  \href {http://adsabs.harvard.edu/abs/2018ApJ...867..153C} {867, 153}

\bibitem[\protect\citeauthoryear{{Costa}, {Sijacki}, {Trenti}  \&
  {Haehnelt}}{{Costa} et~al.}{2014}]{Costa:14}
{Costa} T.,  {Sijacki} D.,  {Trenti} M.,   {Haehnelt} M.~G.,  2014, \mn@doi
  [\mnras] {10.1093/mnras/stu101}, \href
  {https://ui.adsabs.harvard.edu/abs/2014MNRAS.439.2146C} {439, 2146}

\bibitem[\protect\citeauthoryear{{Costa}, {Sijacki}  \& {Haehnelt}}{{Costa}
  et~al.}{2015}]{Costa:15}
{Costa} T.,  {Sijacki} D.,   {Haehnelt} M.~G.,  2015, \mn@doi [\mnras]
  {10.1093/mnrasl/slu193}, \href
  {https://ui.adsabs.harvard.edu/abs/2015MNRAS.448L..30C} {448, L30}

\bibitem[\protect\citeauthoryear{{Costa}, {Rosdahl}, {Sijacki}  \&
  {Haehnelt}}{{Costa} et~al.}{2018}]{Costa:18}
{Costa} T.,  {Rosdahl} J.,  {Sijacki} D.,   {Haehnelt} M.~G.,  2018, \mn@doi
  [\mnras] {10.1093/mnras/sty1514}, \href
  {http://adsabs.harvard.edu/abs/2018MNRAS.479.2079C} {479, 2079}

\bibitem[\protect\citeauthoryear{{Decarli} et~al.,}{{Decarli}
  et~al.}{2017}]{Decarli:17}
{Decarli} R.,  et~al., 2017, \mn@doi [\nat] {10.1038/nature22358}, \href
  {https://ui.adsabs.harvard.edu/abs/2017Natur.545..457D} {545, 457}

\bibitem[\protect\citeauthoryear{{Di Matteo}, {Khandai}, {DeGraf}, {Feng},
  {Croft}, {Lopez}  \& {Springel}}{{Di Matteo} et~al.}{2012}]{DiMatteo:12}
{Di Matteo} T.,  {Khandai} N.,  {DeGraf} C.,  {Feng} Y.,  {Croft} R.~A.~C.,
  {Lopez} J.,   {Springel} V.,  2012, \mn@doi [\apj]
  {10.1088/2041-8205/745/2/L29}, \href
  {https://ui.adsabs.harvard.edu/\#abs/2012ApJ...745L..29D} {745, L29}

\bibitem[\protect\citeauthoryear{{Efstathiou} \& {Rees}}{{Efstathiou} \&
  {Rees}}{1988}]{Efstathiou:88}
{Efstathiou} G.,  {Rees} M.~J.,  1988, \mn@doi [\mnras]
  {10.1093/mnras/230.1.5P}, \href
  {https://ui.adsabs.harvard.edu/\#abs/1988MNRAS.230P...5E} {230, 5p}

\bibitem[\protect\citeauthoryear{{Fan} et~al.,}{{Fan} et~al.}{2001}]{Fan:01}
{Fan} X.,  et~al., 2001, \mn@doi [\aj] {10.1086/324111}, \href
  {https://ui.adsabs.harvard.edu/\#abs/2001AJ....122.2833F} {122, 2833}

\bibitem[\protect\citeauthoryear{{Faucher-Gigu{\`e}re}, {Lidz}, {Zaldarriaga}
  \& {Hernquist}}{{Faucher-Gigu{\`e}re} et~al.}{2009}]{Faucher-Giguere:09}
{Faucher-Gigu{\`e}re} C.-A.,  {Lidz} A.,  {Zaldarriaga} M.,   {Hernquist} L.,
  2009, \mn@doi [\apj] {10.1088/0004-637X/703/2/1416}, \href
  {http://adsabs.harvard.edu/abs/2009ApJ...703.1416F} {703, 1416}

\bibitem[\protect\citeauthoryear{{Habouzit}, {Volonteri}, {Somerville},
  {Dubois}, {Peirani}, {Pichon}  \& {Devriendt}}{{Habouzit}
  et~al.}{2019}]{Habouzit:19}
{Habouzit} M.,  {Volonteri} M.,  {Somerville} R.~S.,  {Dubois} Y.,  {Peirani}
  S.,  {Pichon} C.,   {Devriendt} J.,  2019, \mn@doi [\mnras]
  {10.1093/mnras/stz2105}, \href
  {https://ui.adsabs.harvard.edu/abs/2019MNRAS.tmp.2120H} {p.~2120}

\bibitem[\protect\citeauthoryear{{Hopkins}, {Grudic}, {Wetzel}, {Keres},
  {Gaucher-Giguere}, {Ma}, {Murray}  \& {Butcher}}{{Hopkins}
  et~al.}{2018}]{Hopkins:18}
{Hopkins} P.~F.,  {Grudic} M.~Y.,  {Wetzel} A.~R.,  {Keres} D.,
  {Gaucher-Giguere} C.-A.,  {Ma} X.,  {Murray} N.,   {Butcher} N.,  2018,
  \mnras \, (submitted), \href
  {http://adsabs.harvard.edu/abs/2018arXiv181112462H} {}

\bibitem[\protect\citeauthoryear{{Kannan}, {Marinacci}, {Simpson}, {Glover}  \&
  {Hernquist}}{{Kannan} et~al.}{2018}]{Kannan:18}
{Kannan} R.,  {Marinacci} F.,  {Simpson} C.~M.,  {Glover} S.~C.~O.,
  {Hernquist} L.,  2018, \mnras \, (submitted), \href
  {http://adsabs.harvard.edu/abs/2018arXiv181201614K} {}

\bibitem[\protect\citeauthoryear{{Katz} et~al.,}{{Katz} et~al.}{2019}]{Katz:19}
{Katz} H.,  et~al., 2019, \mnras \, (submitted), \href
  {https://ui.adsabs.harvard.edu/abs/2019arXiv190511414K} {p. arXiv:1905.11414}

\bibitem[\protect\citeauthoryear{{Kim} et~al.,}{{Kim} et~al.}{2009}]{Kim:09}
{Kim} S.,  et~al., 2009, \mn@doi [\apj] {10.1088/0004-637X/695/2/809}, \href
  {http://adsabs.harvard.edu/abs/2009ApJ...695..809K} {695, 809}

\bibitem[\protect\citeauthoryear{{Kimm}, {Cen}, {Devriendt}, {Dubois}  \&
  {Slyz}}{{Kimm} et~al.}{2015}]{Kimm:15}
{Kimm} T.,  {Cen} R.,  {Devriendt} J.,  {Dubois} Y.,   {Slyz} A.,  2015,
  \mn@doi [\mnras] {10.1093/mnras/stv1211}, \href
  {http://adsabs.harvard.edu/abs/2015MNRAS.451.2900K} {451, 2900}

\bibitem[\protect\citeauthoryear{{Kimm}, {Katz}, {Haehnelt}, {Rosdahl},
  {Devriendt}  \& {Slyz}}{{Kimm} et~al.}{2017}]{Kimm:17}
{Kimm} T.,  {Katz} H.,  {Haehnelt} M.,  {Rosdahl} J.,  {Devriendt} J.,   {Slyz}
  A.,  2017, \mn@doi [\mnras] {10.1093/mnras/stx052}, \href
  {http://adsabs.harvard.edu/abs/2017MNRAS.466.4826K} {466, 4826}

\bibitem[\protect\citeauthoryear{{Kimm}, {Haehnelt}, {Blaizot}, {Katz},
  {Michel-Dansac}, {Garel}, {Rosdahl}  \& {Teyssier}}{{Kimm}
  et~al.}{2018}]{Kimm:18}
{Kimm} T.,  {Haehnelt} M.,  {Blaizot} J.,  {Katz} H.,  {Michel-Dansac} L.,
  {Garel} T.,  {Rosdahl} J.,   {Teyssier} R.,  2018, \mn@doi [\mnras]
  {10.1093/mnras/sty126}, \href
  {http://adsabs.harvard.edu/abs/2018MNRAS.475.4617K} {475, 4617}

\bibitem[\protect\citeauthoryear{{Lupi}, {Volonteri}, {Decarli}, {Bovino},
  {Silk}  \& {Bergeron}}{{Lupi} et~al.}{2019}]{Lupi:19}
{Lupi} A.,  {Volonteri} M.,  {Decarli} R.,  {Bovino} S.,  {Silk} J.,
  {Bergeron} J.,  2019, \mn@doi [\mnras] {10.1093/mnras/stz1959}, \href
  {https://ui.adsabs.harvard.edu/abs/2019MNRAS.488.4004L} {488, 4004}

\bibitem[\protect\citeauthoryear{{Mandelker}, {Dekel}, {Ceverino}, {DeGraf},
  {Guo}  \& {Primack}}{{Mandelker} et~al.}{2017}]{Mandelker:17}
{Mandelker} N.,  {Dekel} A.,  {Ceverino} D.,  {DeGraf} C.,  {Guo} Y.,
  {Primack} J.,  2017, \mn@doi [\mnras] {10.1093/mnras/stw2358}, \href
  {http://adsabs.harvard.edu/abs/2017MNRAS.464..635M} {464, 635}

\bibitem[\protect\citeauthoryear{{Moster}, {Naab}  \& {White}}{{Moster}
  et~al.}{2018}]{Moster:18}
{Moster} B.~P.,  {Naab} T.,   {White} S. D.~M.,  2018, \mn@doi [\mnras]
  {10.1093/mnras/sty655}, \href
  {https://ui.adsabs.harvard.edu/abs/2018MNRAS.477.1822M} {477, 1822}

\bibitem[\protect\citeauthoryear{{Peters} et~al.,}{{Peters}
  et~al.}{2017}]{Peters:17}
{Peters} T.,  et~al., 2017, \mn@doi [\mnras] {10.1093/mnras/stw3216}, \href
  {https://ui.adsabs.harvard.edu/abs/2017MNRAS.466.3293P} {466, 3293}

\bibitem[\protect\citeauthoryear{{Rosdahl} \& {Teyssier}}{{Rosdahl} \&
  {Teyssier}}{2015}]{Rosdahl:15b}
{Rosdahl} J.,  {Teyssier} R.,  2015, \mn@doi [\mnras] {10.1093/mnras/stv567},
  \href {http://adsabs.harvard.edu/abs/2015MNRAS.449.4380R} {449, 4380}

\bibitem[\protect\citeauthoryear{{Rosdahl}, {Blaizot}, {Aubert}, {Stranex}  \&
  {Teyssier}}{{Rosdahl} et~al.}{2013}]{Rosdahl:13}
{Rosdahl} J.,  {Blaizot} J.,  {Aubert} D.,  {Stranex} T.,   {Teyssier} R.,
  2013, \mn@doi [\mnras] {10.1093/mnras/stt1722}, \href
  {http://adsabs.harvard.edu/abs/2013MNRAS.436.2188R} {436, 2188}

\bibitem[\protect\citeauthoryear{{Rosdahl}, {Schaye}, {Teyssier}  \&
  {Agertz}}{{Rosdahl} et~al.}{2015}]{Rosdahl:15}
{Rosdahl} J.,  {Schaye} J.,  {Teyssier} R.,   {Agertz} O.,  2015, \mn@doi
  [\mnras] {10.1093/mnras/stv937}, \href
  {http://adsabs.harvard.edu/abs/2015MNRAS.451...34R} {451, 34}

\bibitem[\protect\citeauthoryear{{Rosdahl} et~al.,}{{Rosdahl}
  et~al.}{2018}]{Rosdahl:18}
{Rosdahl} J.,  et~al., 2018, \mn@doi [\mnras] {10.1093/mnras/sty1655}, \href
  {https://ui.adsabs.harvard.edu/abs/2018MNRAS.479..994R} {479, 994}

\bibitem[\protect\citeauthoryear{{Sijacki}, {Springel}  \&
  {Haehnelt}}{{Sijacki} et~al.}{2009}]{Sijacki:09}
{Sijacki} D.,  {Springel} V.,   {Haehnelt} M.~G.,  2009, \mn@doi [\mnras]
  {10.1111/j.1365-2966.2009.15452.x}, \href
  {https://ui.adsabs.harvard.edu/\#abs/2009MNRAS.400..100S} {400, 100}

\bibitem[\protect\citeauthoryear{{Teyssier}}{{Teyssier}}{2002}]{Teyssier:02}
{Teyssier} R.,  2002, \mn@doi [\aap] {10.1051/0004-6361:20011817}, \href
  {https://ui.adsabs.harvard.edu/abs/2002A&A...385..337T} {385, 337}

\bibitem[\protect\citeauthoryear{{Trakhtenbrot}, {Lira}, {Netzer}, {Cicone},
  {Maiolino}  \& {Shemmer}}{{Trakhtenbrot} et~al.}{2017}]{Trakhtenbrot:17}
{Trakhtenbrot} B.,  {Lira} P.,  {Netzer} H.,  {Cicone} C.,  {Maiolino} R.,
  {Shemmer} O.,  2017, \mn@doi [\apj] {10.3847/1538-4357/836/1/8}, \href
  {https://ui.adsabs.harvard.edu/\#abs/2017ApJ...836....8T} {836, 8}

\bibitem[\protect\citeauthoryear{{Tweed}, {Devriendt}, {Blaizot}, {Colombi}  \&
  {Slyz}}{{Tweed} et~al.}{2009}]{Tweed:09}
{Tweed} D.,  {Devriendt} J.,  {Blaizot} J.,  {Colombi} S.,   {Slyz} A.,  2009,
  \mn@doi [\aap] {10.1051/0004-6361/200911787}, \href
  {https://ui.adsabs.harvard.edu/abs/2009A&A...506..647T} {506, 647}

\bibitem[\protect\citeauthoryear{{Volonteri} \& {Rees}}{{Volonteri} \&
  {Rees}}{2006}]{Volonteri:06}
{Volonteri} M.,  {Rees} M.~J.,  2006, \mn@doi [\apj] {10.1086/507444}, \href
  {http://adsabs.harvard.edu/abs/2006ApJ...650..669V} {650, 669}

\bibitem[\protect\citeauthoryear{{Wu} et~al.,}{{Wu} et~al.}{2015}]{Wu:15}
{Wu} X.-B.,  et~al., 2015, \mn@doi [\nat] {10.1038/nature14241}, \href
  {https://ui.adsabs.harvard.edu/\#abs/2015Natur.518..512W} {518, 512}

\makeatother
\end{thebibliography}

%%%%%%%%%%%%%%%%%%%%%%%%%%%%%%%%%%%%%%%%%%%%%%%%%%

% Don't change these lines
\bsp	% typesetting comment
\label{lastpage}
\end{document}